\documentclass[aps,prc,twocolumn,showpacs,preprintnumbers,amsmath,amssymb]{revtex4}

\usepackage{graphicx}
\usepackage{dcolumn}
\usepackage{bm}

\begin{document}

\title{Complex Energy Method in four Body Faddeev-Yakubovsky Equations}

\author{E. Uzu}
\affiliation{Department of Physics, Faculty of Science and Technology,
Tokyo University of Science,
2641 Yamazaki, Noda, Chiba 278-8510, Japan}
\email{j-uzu@ed.noda.tus.ac.jp}
\author{H. Kamada}
\affiliation{Department of Physics, Faculty of Engineering,
Kyushu Institute of Technology,
1-1 Sensuicho, Tobata, Kitakyushu 804-8550, Japan}
\author{Y. Koike}
\affiliation{Science Research Center, Hosei University,
2-17-1 Fujimi, Chiyoda-ku, Tokyo 102-8160, Japan}
\affiliation{Center for Nuclear Study, University of Tokyo,
2-1 Hirosawa, Wako, Saitama, 351-0198, Japan}

\date{\today}

\begin{abstract}
The Complex Energy Method [Prog. Theor. Phys. {\bf 109}, 869L (2003)]
is applied to the four body Faddeev-Yakubovsky equations in  the four nucleon system.
We obtain a well converged solution in all energy regions
below and above the four nucleon break-up threshold.
\end{abstract}


\maketitle

\begin{figure}
\includegraphics[width=65mm]{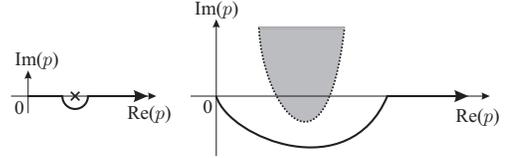}
\caption{Illustration of integration paths
for PVR (left) and CD (right).
The cross  in the left figure indicates a fixed pole.
Moving singularities occur in the shaded area of the right figure.
\label{contour}}
\end{figure}

Calculations for scattering systems in configuration space require boundary conditions
which increase in complexity with growing particle numbers.
These boundary conditions appear in the form 
of Green's functions in momentum space
which carry singularities of increasing complexity.
The Green's functions are expressed as $G_0 = 1 / (E + i \varepsilon - H_0)$
where $E$ and $H_0$ are the  total and kinetic energy, respectively,
and the limit $\varepsilon \to 0$ has to be taken. 
In the two-body system there is one (relative) momentum variable $p$ and $G_0$
has a pole in the complex $p$-plane. 
It is easy to handle it using the principal value prescription and (half)
the residue theorem (PVR). 
In the three body case there arises already a difficulty in the form of
so called moving singularities \cite{Schmid,HKWG},
however, PVR is still applicable~\cite{GWHKG},
or one can use the contour deformation~\cite{LVLC,CaSl,Eben,Koike} (CD) technique.
Summarizing these techniques, first one takes the  limiting value  $\varepsilon \to 0$
and next the equation is solved avoiding the integration path on the complex plane.
This is illustrated in Fig. \ref{contour}.

\begin{figure}
\includegraphics[width=50mm]{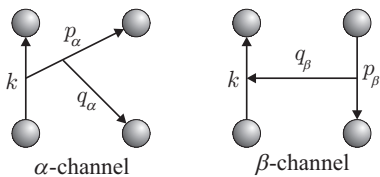}
\caption{The two partitions in  four particle system.
$k$, $p$'s, and $q$'s are standard Jacobi momenta.
\label{channel}}
\end{figure}

The situation is more complicated in the four body system.
Employing a separable potential and a separable expansion technique
for  the three body, and [2+2] subamplitudes, the four body
Faddeev-Yakubovsky (FY) equations~\cite{FaYa} 
for four identical particles can  be  expressed as
\begin{widetext}
\begin{equation}
\begin{pmatrix}
{\cal M}_{\alpha \alpha} & {\cal M}_{\alpha \beta} \cr
{\cal M}_{\beta \alpha}  & {\cal M}_{\beta \beta}
\end{pmatrix} =
\begin{pmatrix}
\pm {\cal E} & {\cal F}_1 \cr
2 ({\cal F}_1^T + {\cal F}_2^T) & 0
\end{pmatrix} +
\begin{pmatrix}
\pm {\cal E} & {\cal F}_1 \cr
2 ({\cal F}_1^T + {\cal F}_2^T) & 0
\end{pmatrix}
\begin{pmatrix}
{\cal H} & 0 \cr 0 & {\cal G}
\end{pmatrix}
\begin{pmatrix}
{\cal M}_{\alpha \alpha} & {\cal M}_{\alpha \beta} \cr
{\cal M}_{\beta \alpha}  & {\cal M}_{\beta \beta}
\end{pmatrix}, \label{FYeq}
\end{equation}
\end{widetext}
where the ${\cal M}$'s are the four body amplitudes,
$\alpha$ and $\beta$ indicate [3+1] and [2+2] configurations (see Fig. \ref{channel}),
${\cal E}$ is an exchange term from [3+1] to [3+1] configurations
of which the plus sign corresponds to the four bosons
and minus sign to the fermions.
${\cal F}$'s are exchange terms from [2+2] to [3+1]
where the subscripts are related to the two diagrams in Fig. 3.
More details may be found in Ref.~\cite{OKSNS}.
Further ${\cal H}$ and ${\cal G}$ are the three body and [2+2] propagators
and they have a similar nature as Green's function
in the two body Lippmann-Schwinger (LS) equation.

If one stays below the three-body break-up threshold the FY equations 
can be solved with PVR, since only two-body singularities occur
\cite{uzu1,uzu2,uzu3,FON1,FON2,FON3,FON4}.
Above the three-body threshold but still below the four-body threshold 
the FY equations have also been solved applying the CD techniques
~\cite{FON1,FON2,FON3,FON4,FON5,FON6}.
There in the $\cal E$ and $\cal F$'s terms  occur two-body propagators whose nature 
is similar to the three body Green's function
in the Born term of the three body Alt-Grassberger-Sandhas~\cite{AGS} or
Amado-Mitra-Faddeev-Lovelace (e.g. \cite{AMFL}) equations.
However, above the four-body threshold the four-body Green's function 
depends on all (relative) momenta  and the behavior of those singularities 
is quite complicated. 
Thus neither PVR nor CD techniques have  been successfully extended
at energies above the four body break-up threshold
and we are not aware of a solution in this energy region.

\begin{figure}
\includegraphics[width=70mm]{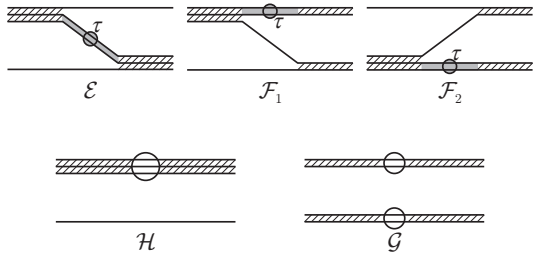}
\caption{Diagrams for the ${\cal E}$, ${\cal F}_1$, ${\cal F}_2$, ${\cal H}$, and ${\cal G}$ 
ingredients to Eq. (1).  
The shaded parts with $\tau$ indicate the two body propagators.
\label{diagram}}
\end{figure}

Recently the Complex Energy Method~\cite{CEM} (CEM) has been revived 
and applied to the two- and three-nucleon system.
The first step of CEM is to solve the equation with some finite $\varepsilon$'s.
These calculations are easily carried out since there are no singularities on the real momentum axis.
After obtaining solutions with various $\varepsilon$'s,
the limiting value  $\varepsilon \to 0$ is taken numerically
with an analytical continuation method.

Our aim is to generate  solutions  applying this method to the FY equations
in all energy regions including energies  above the four body break-up threshold.
We performed calculations in the  $J^{\pi} = 0^+$ and $T=0$ state for the four nucleon system.
For this  feasibility  study  the  $J^{\pi} = 1/2^+$ state is included in the three body subsystem
and the $^1$S$_0$ and $^3$S$_1$-$^3$D$_1$ states in the two body subsystem.
All allowed spins and angular momenta within this restriction are included ;
in short, there are 14 channels.
The Coulomb force is neglected.
The Yamaguchi potential~\cite{YY,ACP} is employed as the nucleon-nucleon interaction.
The potential $V$ has a separable form as
\begin{equation}
V_{\ell \ell' }(k,k') = g_{\ell}(k) \lambda g_{\ell'}(k'),
\end{equation}
where the $g_\ell$'s are the two body form factors for the partial waves $\ell$ as
\begin{equation}
g_0(k) = \frac{\, 1 \,}{\, k^2 + \beta^2 \,}, ~~~~
g_2(k) = \frac{\, C k^2 \,}{\, ( k^2 + \beta^2 )^2 \,},
\end{equation}
and the  $k$'s are the initial and final momenta between the two nucleons, respectively.
We adopt the parameters $\lambda$, $\beta$, and $C$ given in  Table~\ref{tbl1}.
We represent the three body and [2+2] subamplitudes by  rank-4 separable forms
employing the Energy Dependent Pole Expansion~\cite{EDPE} (EDPE) method.
We take the nucleon mass as  938.91897MeV which is the average
of those for proton and neutron,
and $\hbar c = 197.327054$ MeV$\cdot$fm.
The integrations are cut off at 200 fm$^{-1}$ for $k$, at 
40 fm$^{-1}$ for the $p$'s, and at 16 fm$^{-1}$ for the $q$'s (see Fig. \ref{diagram}).

\begin{table}[b]
\caption{Parameters of the Yamaguchi potential.
\label{tbl1}}
\begin{ruledtabular}
\begin{tabular}{ll|lll}
state               & $\lambda$ (MeV fm$^{-1}$) & state & $\beta$ (fm$^{-1}$) & C (fm$^{0}$) \\ \hline
$^1$S$_0$           & -68.942626 & $^1$S$_0$ & 1.1300 &            \\
$^3$S$_1$-$^3$D$_1$ & -74.506955 & $^3$S$_1$ & 1.2412 &            \\
                    &            & $^3$D$_1$ & 1.9476 & -4.4950154 \\
\end{tabular}
\end{ruledtabular}
\end{table}

\begin{figure*}
\includegraphics[width=140mm]{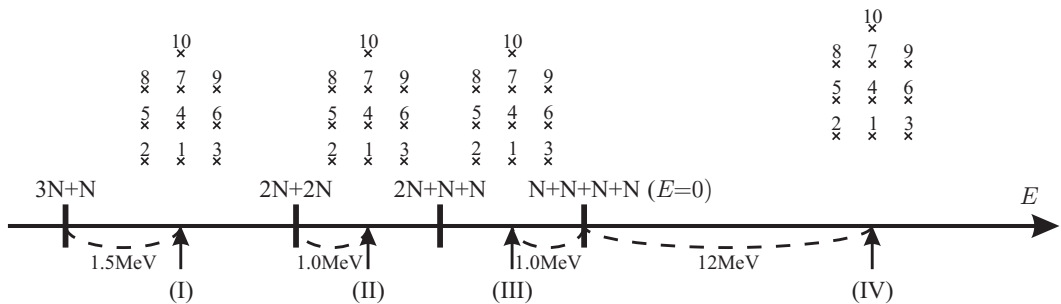}
\caption{Illustration of threshold energies for the 4N system.
We choose E=0 at the four-body threshold. The various energies for the calculations 
are measure relative to the thresholds.
The crosses indicate the complex energies where we solve the FY equations in the CEM.
They are numbered by  $n$ for each choice of the energy region (I-IV).
\label{thresh}}
\end{figure*}

The FY equations are solved at four energies: (I) 1.5MeV above the 3N+N threshold,
(II) 1MeV above the 2N+2N threshold, (III) 1MeV below the four body break-up threshold,
and (IV) 12MeV above  it (see Fig.~\ref{thresh}).
We define the $i \varepsilon$ term of the four body Green's function as
$i \varepsilon + \zeta$,  where $\varepsilon$ and $\zeta$ are real. 
Thus $G_0$ turns into $G_0 =1/(E+i\varepsilon+\zeta-H_0)$.
Solutions of the FY equation satisfy uniqueness
even at the limit for $\varepsilon \to 0$,
which is not the case for simple LS equation~\cite{gloeckle}.
Therefore the results by the analytical continuation do not
depend on the choice of $\varepsilon$'s within the radius of convergence.
Thus we empirically choose 0.5MeV as the minimum $\varepsilon$ value
for the cases (I)-(III) and 0.75MeV for the case (IV)
(see crosses in Fig.~\ref{thresh}), with attention only to a beter numerics.
They are increased in steps of 0.125MeV.
$\zeta$ is chosen as  0 and $\pm$0.125MeV.

We employ the point method~\cite{PM} as an analytical continuation technique in CEM.
It's convergence behavior is shown in Table \ref{tab2}
where  the phase shift $\delta$ and the inelasticity parameter $\eta$
is defined by $S = \eta \exp (2 i \delta)$.  Here $S$ is the S-matrix
of elastic 3N+N scattering and is related 
to the on-shell amplitude of ${\cal M}_{\alpha \alpha}$
in Eq.~(\ref{FYeq}) (${\cal M}_{\alpha \alpha}^{\rm on}$) as
$S = 1 - 2 i \kappa {\cal M}_{\alpha \alpha}^{\rm on}$
where $\kappa$ is the on-shell momentum.

In case (I) $\eta$ must be 1 due to unitarity
and our result satisfies it within six digits.
Also in cases (II) and (III) we reach a very high accuracy. 
In case (IV), we still obtain converged solutions within  4 digits.
In the cases (I and II) our results agree very well with the solutions 
based on PVR.

\begin{table*}
\caption{Phase shifts (in deg) and inelasticity parameters
for 3N+N $\to$ 3N+N elastic scattering.
$n_{\rm max}$ denotes the number of sample energies which  are included  in the point method.
For instance, $n_{\rm max} = 5$ means that the solutions from $n$=1 to  5
(see Fig.~\ref{thresh})  are included.
The row PVR shows results from  a direct solution of the FY equations
 using PVR. The agreement is perfect.
\label{tab2}}
\begin{ruledtabular}
\begin{tabular}{r|cc|cc|cc|cc}
& \multicolumn{2}{c|}{(I)} & \multicolumn{2}{c|}{(II)}
& \multicolumn{2}{c|}{(III)} & \multicolumn{2}{c}{(IV)} \\ \hline
$n_{\rm max}$ &$\delta$ & $\eta$ & $\delta$ & $\eta$ &
     $\delta$ & $\eta$ & $\delta$ & $\eta$ \\ \hline
  1 & 53.31351 & 1.018810 & 14.62234 & 0.675088 & -8.0617 & 0.692391 & -62.093 & 0.83875 \\
  2 & 46.21307 & 0.973139 & 10.47942 & 0.853658 & -5.9428 & 0.815698 & -61.965 & 0.75946 \\
  3 & 44.27898 & 0.989232 & 12.38204 & 0.948787 & -5.5022 & 0.899150 & -61.620 & 0.74499 \\
  4 & 44.27129 & 1.000623 & 12.38254 & 0.948044 & -5.5101 & 0.898666 & -61.676 & 0.74570 \\
  5 & 44.34441 & 0.999787 & 12.38211 & 0.948046 & -5.5095 & 0.898656 & -61.682 & 0.74589 \\
  6 & 44.34157 & 0.999994 & 12.38284 & 0.948053 & -5.5094 & 0.898655 & -61.669 & 0.74580 \\
  7 & 44.34022 & 1.000005 & 12.38198 & 0.948070 & -5.5096 & 0.898654 & -61.666 & 0.74570 \\
  8 & 44.34012 & 0.999997 & 12.38198 & 0.948069 & -5.5095 & 0.898657 & -61.669 & 0.74581 \\
  9 & 44.34013 & 0.999999 & 12.38198 & 0.948069 & -5.5096 & 0.898656 & -61.670 & 0.74580 \\
 10 & 44.34016 & 1.000000 & 12.38198 & 0.948069 & -5.5095 & 0.898657 & -61.669 & 0.74582 \\ \hline
PVR & 44.34016 & 1.000000 & 12.38198 & 0.948069 & ------  &  ------  & ------  &  ------ \\ 
\end{tabular}
\end{ruledtabular}
\end{table*}

We showed that well converged solutions of the FY equations are
obtained in all energy regions.
In relation to the application of EDPE we confirmed that converged solutions
are obtained in the cases (I-III). In the case (IV), however, there is a report
that EDPE is not applicable~\cite{POS}.
We also found that EDPE did not converge. Therefore, in this 
study we just kept the rank fixed by 4.
We plan to investigate  this problem in a 
forthcoming study.

Further we shall include higher partial waves and employ realistic NN forces to 
discuss physics. One expects that evidence for three-nucleon forces is more 
pronounced in the high energy region and the presented method is applicable 
there, now in the four-nucleon system.

\section*{Acknowlegements}

Authors would like to thank Prof. W. Gl\"ockle for helping 
us to read this manuscript carefully.
The calculations are performed on SX-5/128M8 (Research Center
for Nuclear Physics), SX-5/6B (National Institute for
Fusion Science), HP X4000 (Frontier Research Center for
Computational Science, Tokyo University of Science) in Japan,
and partly on Hitachi SR8000 (Leibnitz-Rechenzentrum
f\"ur die M\"unchener Hochschule) in Germany.

\end{document}